\documentclass[twocolumn,showpacs,preprintnumbers,amsmath,amssymb]{revtex4}

\usepackage{graphicx}
\usepackage{dcolumn} 
\usepackage{bm}      
\usepackage{color}

\usepackage{textcomp}

\newcommand{\LN} {$\rm LiNbO_3$ }

\begin{document}

\preprint{APS/123-QED}

\title{Impact of the tip radius on the lateral resolution\\ in piezoresponse force microscopy}

\author{Tobias Jungk}
\email{jungk@uni-bonn.de}
\affiliation{Institute of Physics,
University of Bonn, Wegelerstra\ss e 8, 53115 Bonn, Germany\\}
\author{\'{A}kos Hoffmann}
\affiliation{Institute of Physics, University of Bonn,
Wegelerstra\ss e 8, 53115 Bonn, Germany\\}
\author{Elisabeth Soergel}
\affiliation{Institute of Physics, University of Bonn,
Wegelerstra\ss e 8, 53115 Bonn, Germany\\}

\date{\today}

\begin{abstract}
We present a quantitative investigation of the impact of tip radius
as well as sample type and thickness on the lateral resolution in
piezoresponse force microscopy (PFM) investigating bulk single
crystals.
The observed linear dependence of the width of the domain wall on
the tip radius as well as the independence of the lateral resolution
on the specific crystal-type are validated by a simple theoretical
model.
Using a Ti-Pt-coated tip with a nominal radius of 15\,nm the so far
highest lateral resolution in bulk crystals of only 17\,nm  was
obtained.
\end{abstract}


\maketitle

Ferroelectrics attract increasing attention due to their
applicability, e.g., for electrically controlled optical
elements~\cite{Eas01}, for efficient
frequency-doubling~\cite{Fej92,Ilc04}, for photonic
crystals~\cite{Bro00}, or for non-volatile memories with an
otherwise un-reached data-storage density~\cite{Cho05a}.
The size of the domain structures required for those applications
varies from a few microns down to some nanometers.
The smaller the domains are the more the properties of the domain
walls become important.
Generally the width of the domain walls is expected to be a few
crystal lattice cells~\cite{Pad96,Mey02,Cat07}.
Measuring the birefringence with scanning near-field microscopy in
$\rm LiTaO_3$, however, showed a distortion of the crystal over a
width of 3\,\textmu m across the domain boundary~\cite{Yan99}.
Detecting the Raman modes with a confocal defect-luminescence
microscope, an influence of the domain wall in \LN further than
1\,\textmu m into the surrounding material was observed
\cite{Die03}.
On the other hand, high-resolution transmission electron microscopy
yielded a domain wall width in PbTiO$_3$ of $<2.5$\,nm~\cite{Foe99}
and with scanning nonlinear dielectric microscopy, a width of only
0.5\,nm in ultra-thin PZT films was determined~\cite{Cho05b}.

In the past ten years, piezoresponse force microscopy (PFM) has
become a very common technique for domain imaging mainly due to its
high lateral resolution without any need for specific sample
preparation. In brief, for PFM a scanning force microscope is
operated in contact mode with an alternating voltage applied to the
tip. In ferroelectric samples this voltage causes thickness changes
via the converse piezoelectric effect~\cite{New} and therefore
vibrations of the surface which lead to oscillations of the
cantilever that can be read out with a lock-in
amplifier~\cite{Kol95,Alexe}.

In this contribution, we present a study of the influence of the tip
radius on the PFM imaging of 180$^{\circ}$ ferroelectric domain
walls in bulk single crystals. Therefore we determined the width $W$
of domain walls when imaged by PFM on \LN crystals for tips of
different radii. We also compared $W$ for different types of
crystals using one unique tip. Furthermore, a series of measurements
was carried out with \LN of different sample thicknesses. Finally we
present a simple analytical model that explains the observed
dependencies of the domain wall $W$ on the tip radius, the crystal
thickness and type of sample.

The experiments were carried out with two different scanning force
microscopes (Topometrix Explorer from Veeco and SMENA from NT-MDT).
Both systems were modi\-fied to allow application of voltages to the
tip in order to enable PFM measurements. The alternating voltage
($U_{\rm tip}=10\rm\,V_{pp}, 30-60$\,kHz) was applied to the tip and
the backside of the samples was grounded. To obtain reliable data
for the width $W$ we readout the $X$-signal of the lock-in amplifier
(in the following denoted as PFM signal). The recorded data were
thus unaffected by the background inherent to PFM
measurements~\cite{Jun06} which can lead to a presumed broadening of
the width W~\cite{Jun07}.

We used a series of different tips with radii varying from
$10-90$\,nm, classified in Tab.~\ref{tab:Jungk1}. All tips were made
out of highly n-doped silicon and conductively coated with different
materials. The spring constants ranging from 3 to 70 \,N/m. The tip
with the smallest radius ($r = 10$\,nm) makes an exception since it
was uncoated. Due to oxidization the outer few nanometers of the
surface are modified to non-conductive SiO$_2$. As a consequence,
the mechanical and the electrical tip do not coincide any more; the
SiO$_2$-layer acts as a dielectric gap between tip and sample.
\begin{table}
\caption{\label{tab:Jungk1}
Specifications of the different tips utilized for the measurements.}
\begin{ruledtabular}
\begin{tabular}{crclcc}
Label & Manufacturer& &Model  & Tip          & Coating   \\
      &             & &       & radius [nm]  &           \\
\hline
A & NT-MDT     \footnotemark[1] &&NSG11     & 10 & --    \\
B & Veeco      \footnotemark[2] &&OSCM-PT   & 15 & Ti-Pt \\
C & Veeco      \footnotemark[2] &&SCM-PIT   & 20 & Pt-Ir \\
D & MikroMash  \footnotemark[3] &&NSC35     & 35 & Ti-Pt \\
E & MikroMash  \footnotemark[3] &&NSC35     & 50 & Cr-Au \\
F & NT-MDT     \footnotemark[1] &&DCP11     & 50--70 & diamond \\
G & MikroMash  \footnotemark[3] &&NSC35     & 90 & Co-Cr \\
\end{tabular}
\end{ruledtabular}
\footnotetext[1]{www.ntmdt.ru} \footnotetext[2]{www.veeco.com}
\footnotetext[3]{www.spmtips.com}
\end{table}

The experiments for determining the dependence of the width  $W$ on
the tip radius $r$ were performed with periodically poled (period
length $\Lambda = 30$\,\textmu m) congruently melting 500\,\textmu m
thick \LN crystals (PPLN).
For the measurements of $W$ on the sample thickness $t$ we used the
same \LN samples, mechanically thinned by polishing to the thickness
wanted  ($15-1000$\,\textmu m). The thinnest sample (0.9\,\textmu m)
was a stoichiometric \LN crystal. Here the domains were generated
with the help of the tip by applying a voltage of 20\,V for 10\,min.
We also measured a series of samples different from \LN as listed in
Tab.~\ref{tab:Jungk2}. Those samples had thicknesses of $0.5-2$\,mm
and were either periodically poled (KTiOPO$_4$ and LiNbO$_3$) or had
arbitrary domain patterns ($\rm BaTiO_3, KNbO_3, LiTaO_3$ and
Sr$_{0.61}$Ba$_{0.39}$Nb$_2$O$_6$).

Calculating the spatial resolution achievable with PFM requires the
exact electric field distribution underneath the tip and the
electromechanical answer of the material. The latter is given by the
dielectric constants as well as the elastic and piezoelectric
tensors, respectively. This complex problem is most suitable for the
finite element method (FEM), where all material constants can be
included thus yielding quantitative results~\cite{All70}. A detailed
study on domain wall width imaged by PFM including FEM-calculations
has recently been undergone~\cite{Tia07}. In this contribution, we
propose a much simplified approach to the problem of lateral
resolution in PFM. This model is not capable of giving the amplitude
of the measured PFM signals because we performed some normalization
in order to facilitate the calculations. The model can, however,
give a quantitative prediction of the width~$W$ as a function of the
tip radius~$r$ and the sample thickness~$t$.

\begin{figure}
\includegraphics{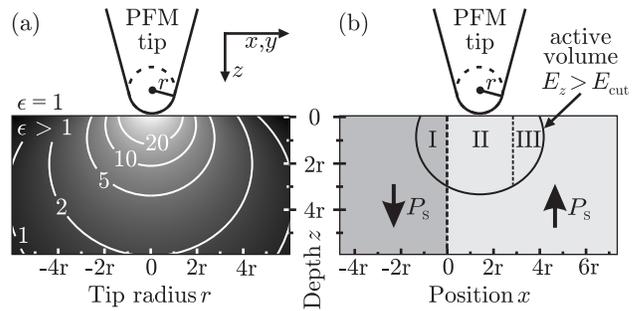}
\caption{\label{fig:Jungk1}
(a) Electrical field distribution $E_z$ underneath a tip of radius
$r$ calculated with Eq.~\ref{eq:Jungk1} for a sample of infinite
thickness. The white lines indicate the shells where the electric
field has decayed to the indicated \%-value of the maximum electric
field $E_{\rm max}$ present underneath the tip apex. (b) Schematics
of the analytical model with the tip in the vicinity of a domain
wall. The active volume where $E_z>E_{\rm cut} \approx 0.05\,E_{\rm
max}$ underneath the tip covers both domains. The contributions of
part I and III cancel out each other. The resulting surface
deformation is thus determined by part II only.
}
\end{figure}

In a first step the problem was simplified by approximating the
spherical apex of the tip with radius $r$  by a point charge at the
distance $r$ above the sample surface. We further assumed the sample
to be isotropic with an effective dielectric constant
$\varepsilon_{\rm eff} = \sqrt{\varepsilon_r \, \varepsilon_z}$,
$\varepsilon_z$ being the dielectric constant in $z$-direction and
$\varepsilon_r$ the radial one perpendicular to $z$. The electric
field distribution underneath the tip is then given
by~\cite{Greiner}:
\begin{eqnarray}\label{eq:Jungk1}
E_{\rm z}(x,y,z) = \frac{q}{\varepsilon_{\rm eff}}
&& \Bigg \{
\frac{z+r}{\left[ x^2+y^2+\left(z+r\right)^2
\right]^{3/2}}\\
&& +  \frac{z-r-2t}{\left[ x^2+y^2+ \left( z-2t-r\right)^2
\right]^{3/2}}
\Bigg\}\,.\nonumber
\end{eqnarray}
Next we define $E_{\rm max}$ as the maximum electric field
underneath the tip apex inside the sample (at $x=y=z=0$). $E_{\rm
max}$ will be used to normalize the electric field strength.
Furthermore, we scale all lengths with the tip radius $r$. The
problem has thus become dimensionless. Figure~\ref{fig:Jungk1}(a)
shows the electric field distribution $E_z$ inside a sample of
infinite thickness. It can be seen, that at a depth of twice the tip
radius $r$, the electric field has decayed to almost 10\% of its
initial value $E_{\rm max}$.

Figure~\ref{fig:Jungk1}(b) illustrates the analytical model used for
the simulation of the lateral resolution of PFM depending on the tip
radius and the sample thickness. First we define an {\sl active
volume} inside of which the electric field has decreased to a value
$E_{\rm cut}$ which has to be identified later by fitting the
experimental data. Beyond the active volume we set $E_z=0$. The
resulting piezomechanical deformation is calculated by integrating
the contributions of the sample within the active volume:
\begin{equation}
\Delta z(x) = d_{33}~z~\int_{-E_{\rm cut}}^{E_{\rm cut}}
\int_{0}^{E_{\rm cut}} E_z (x,y,z)~{\rm d}E_y {\rm d}E_z\,.
\end{equation}
As can be seen in Fig.~\ref{fig:Jungk1}(b) the contributions of
region I and III to the emerging piezomechanical deformation cancel
out each other. For the simulation of the PFM signal when scanning
across a domain wall, the active volume was subdivided into approx.\
10$^9$~cubic elements. In order to determine $E_{\rm cut}$ the
experimental data was fitted with the calculated slopes of the PFM
signal across the domain wall. The best fits were obtained for
$E_{\rm cut} = 0.92 \, E_{\rm max}$ for all tip radii.
Note that the main assumption made for the simulation consists in
the stiffness of the crystal, i.e., within a length of some microns
the crystal parts can not deform independently. The strength of
clamping in bulk samples has been demonstrated by other
experiments~\cite{Jun07a}.


%
\begin{figure}[ttt]
\includegraphics{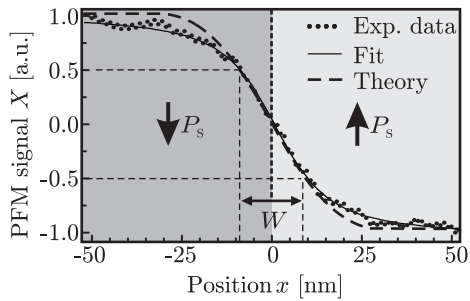}
\caption{\label{fig:Jungk2}
Measured PFM signal line scan ($\bullet$) across a 180$^{\circ}$
domain wall in \LN  recorded with a tip of type B ($r=15$\,nm). The
domain wall width $W$ is determined fitting the data using
Eq.~\ref{eq:Jungk2}. For comparison, a line scan is shown that was
calculated with the theoretical model for a tip with a radius of
15\,nm .
}
\end{figure}

In order to deduce the domain wall width $W$ from the experimental
data we normalized the PFM signals to an amplitude of~1 and fitted
the data with a modified hyperbolic tangent
\begin{equation}
\label{eq:Jungk2}
X(x) = A \tanh\left(\frac{x}{w}\right) + B
\arctan\left(\frac{x}{w}\right)
\end{equation}
where $A$, $B$ and $w$ are used as free parameters. Note that this
function is only used to determine the width $W$ but has no direct
physical meaning. We then applied a $25\% - 75\%$ criterion on the
scan lines of the PFM signal which corresponds to the full width at
half maximum of the PFM amplitude if no PFM background is
present~\cite{Jun07}. Figure~\ref{fig:Jungk2} shows an example for a
$r=15$\,nm tip on a \LN sample. The measurement (dotted line) is
fitted according to Eq.~\ref{eq:Jungk2}  (slim line). For comparison
the slope of the PFM signal calculated with the analytical model is
also depicted (dashed line).
As can be seen, the error determining $W$ using Eq.~\ref{eq:Jungk2}
is minimal that is why we do not show any error bars in the
subsequent graphs.

\begin{figure}[ttt]
\includegraphics{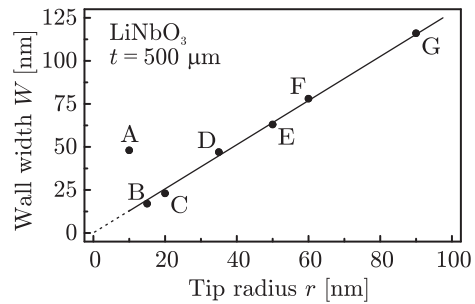}
\caption{\label{fig:Jungk3}
Measured domain wall width $W$ as a function of the nominal tip
radius $r$ (Tab.~\ref{tab:Jungk1}). The straight line was calculated
using the analytical model presented in this contribution. Note that
despite its smaller radius, the uncoated tip A shows an inferior
lateral resolution.
}
\end{figure}

Processing the data in the above described manner, we extracted the
data for $W$ as a function of the tip radius as shown in
Fig.~\ref{fig:Jungk3}. The straight line results from our analytical
model. The excellent agreement between measurement and model is
striking and strongly sustains the model to give reasonable
estimates on $W$. Implicit to our model, an atomically sharp domain
wall is thus also sustained by the PFM measurements. Note that with
tip B a width of only 17\,nm was measured, the smallest value
recorded with PFM in bulk materials so far.
This has to be compared with a recent publication, where a lowest
limit for $W$ of 65\,nm in \LN was estimated~\cite{Rod05}. This
value as well as our currently highest resolution is by no means a
fundamental limit of the material itself but by the available tip
sizes and imaging parameters.
It is furthermore evidently seen that the non-coated tip A shows a
substantially reduced lateral resolution ($W\approx 50\,$nm) than
pretended by its tip radius of only 10\,nm. This, however, is
exactly what can be expected from an surface oxide layer: the
conductive part of the tip being at a distance of some nanometers
from the sample surface, separated by the dielectric SiO$_2$-layer
generates an less localized electrical field inside the crystal,
thus the reduced spatial resolution.

\begin{figure}[bbb]
\includegraphics{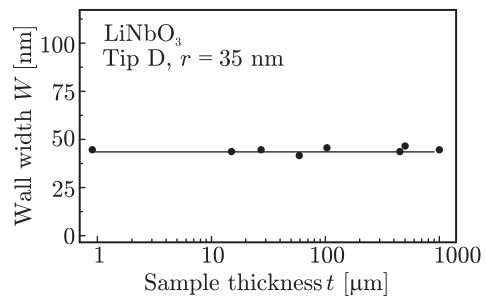}
\caption{\label{fig:Jungk4}
Measured domain wall width $W$ as a function of sample thickness
$t$. The straight line is the result from the analytical model.
}
\end{figure}
In a further series of experiments we determined $W$ as a function
of the sample thickness $t$ varying the latter by three orders of
magnitude from 0.9 -- 1000\,\textmu m (Fig.~\ref{fig:Jungk4}). We
used tips of type D with a nominal radius $r=35$\,nm. The accuracy
and durability of the tips ware controlled by measuring a standard
500\,\textmu m thick PPLN sample before and after each data
acquisition with a sample of modified thickness. From
Fig.~\ref{fig:Jungk4} no change of the width $W$ within the
thicknesses range of the samples could be observed.
This, however, is consistent with our theoretical model where the
electric field distribution $E_z$ is found to be independent on the
sample thickness $t$ for $t>15r$. In the case of tip D with a
nominal radius of $r=35$\,nm the electric field $E_z$ is thus the
same for samples with thicknesses $>500$\,nm.

Finally we compared $W$ for different crystals using tips of type B
and D (Tab.~\ref{tab:Jungk2}). We again checked the reliability of
the recorded data by always recording comparative measurements with
a standard PPLN sample. All samples show the same width $W$ for a
specifice tip of radius $r = 15$\,nm or $r = 35$\,nm within an error
of $\pm 1$\,nm, although their dielectric anisotropy $\gamma =
\sqrt{\varepsilon_z/\varepsilon_{r}}$ differ as listed in
Tab.~\ref{tab:Jungk1}. This can be understood if we calculate the
electrical field e.~g.\ with the method of image charges where we
place the charge $q$ at $(0,0,-r)$. In contrast to the case of an
infinite isotropic half-plane, characterized by a single dielectric
constant, we need two image charges $q'$ at $(0,0,r)$ and
$\tilde{q}$ at $(0,0,\tilde{r})$ to account for the dielectric
anisotropy given by $\varepsilon_z$ and $\varepsilon_r$. The
potential distribution above the surface is identical to the
isotropic case but inside the crystal it is given by
\begin{equation}
\varphi(x,y,z) = \frac{\tilde{q}}{\sqrt{{\varepsilon_r}^2
\varepsilon_z}} \frac{1}{\sqrt{\left(x^2+y^2\right)/\varepsilon_r +
\left(z-\tilde{r}\right)^2/\varepsilon_z}} \,.
\end{equation}
The only way to satisfy the boundary conditions at the surface
$(x,y,0)$ is to set $\tilde{r}=\gamma r$ which leads to
\begin{equation}
E_{\rm z}(x,y,z) = \frac{2q\gamma}{1+\varepsilon_{\rm eff}}
\frac{z+r}{\left[ x^2 + y^2 + \left(z+r \right)^2 \right]^{3/2}}\,.
\end{equation}
From this equation we can clearly see that the dielectric anisotropy
does not affect the shape of the electrical field distribution but
only the field strength.
\begin{table}[ttt]
\caption{\label{tab:Jungk2}
Domain wall width $W$ measured for different samples with tips of
type B ($r=15$\,nm) and D ($r=35$\,nm). c-LiNbO$_3$: congruently
melting and s-LiNbO$_3$: stoichiometric lithium niobate
respectively, SBN: Sr$_{0.61}$Ba$_{0.39}$Nb$_2$O$_6$.}
\begin{ruledtabular}
\begin{tabular}{lccc}
    Sample & Domain wall width   & Dielectric anisotropy                      \\
           & W\,[nm]& $ \gamma = \sqrt{\varepsilon_z / \varepsilon_{r}} $  \\
           & $r=15$\,nm / $r=35$\,nm  \\
\hline
    BaTiO$_3$  & 19~~/~~46 & $0.18$ \\
    KNbO$_3$   & 18~~/~~45 & $0.34$ \\
    KTiOPO$_4$ & 17~~/~~46 & $0.86$ \\
    LiTaO$_3$  & 18~~/~~45 & $0.89$ \\
    c-\LN      & 17~~/~~46 & $0.58$ \\
    s-\LN      & 17~~/~~45 & $0.58$ \\
    Mg:\LN     & 18~~/~~47 & $0.58$ \\
    SBN        & 18~~/~~48 & $1.30$ \\
\end{tabular}
\end{ruledtabular}
\end{table}

To summarize, we have analyzed the impact of the tip radius and the
sample thickness on the domain wall width $W$ observed with
piezoresponse force microscopy.
We introduced an analytical model which explains the experimental
data: a linear dependency of $W$ on the tip radius as well as no
dependency of $W$ on the sample thickness as long as the sample is
thicker than 15-times the tip radius.
The model assuming an infinite sharp domain wall is perfectly
consistent with the experimental data.
Furthermore we carried out a series of measurements for different
kind of single crystals. Even though their material parameters
differ significantly all measured domain wall width were found to be
independent of the specific sample. This is explained by the
identical shape of the electrical field inside the sample which is
the basis of our analytical model.

\noindent{\bf Acknowledgments}

\noindent We thank L.~Tian and V.~Gopalan for fruitful discussions.
\newline
We thank G.~Baldenberger from the Institut National d'Optique
(Canada) for providing the PPLN samples, D.~Rytz from the
Forschungsinstitut f\"{u}r mineralische und metallische Werkstoffe
(Germany) for providing the BaTiO$_3$ and KNbO$_3$ samples.
Financial support of the DFG research unit 557 and of the Deutsche
Telekom AG is gratefully acknowledged.


\begin{thebibliography}{}


\bibitem{Eas01}
 R.~W.~Eason, A.~S.~Boyland, S.~Mailis, and P.~G.~R.~Smith,
Opt. Commun. \textbf{197}, 201(2001).

\bibitem{Fej92}
M.~M.~Fejer, G.~A.~Magel, D.~H.~Jundt, and R.~L.~Byer, IEEE\ J.\
Quantum Electron.\ \textbf{28}, 2631 (1992).

\bibitem{Ilc04}
V.~S.~Ilchenko, A.~A.~Savchenkov, A.~B.~Matsko, and L.~Maleki,
Phys.\ Rev.\ Lett.\ \textbf{92}, 043903 (2004).

\bibitem{Bro00}
N.~G.~R.~Broderick, G.~W.~Ross, H.~L.~Offerhaus, D.~J.~Richardson,
and D.~C.~Hanna, Phys.\ Rev.\ Lett.\ \textbf{84}, 4345 (2000).

\bibitem{Cho05a}
Y.~Cho, S.~Hashimoto, N.~Odagawa, K.~Tanaka, and Y.~Hiranaga,
Appl.\ Phys.\ Lett.\ \textbf{87}, 232907 (2005).

\bibitem{Pad96}
J.~Padilla, W.~Zhong, and D.~Vanderbilt, Phys.\ Rev.\ B \textbf{53},
R5969 (1996).

\bibitem{Mey02}
B.~Meyer and D.~Vanderbilt, Phys.\ Rev.\ B \textbf{65}, 104111
(2002).

\bibitem{Cat07}
G.~Catalan, J.~F.~Scott, A.~Schilling, and J.~M.~Gregg, J.~Phys.:
Condens.\ Matter \textbf{19}, 022201 (2007).

\bibitem{Yan99}
T.~J.~Yang,V.~Gopalan, P.~J.~Swart, and U.~Mohideen, Phys. Rev.
Lett. \textbf{82}, 4106 (1999).

\bibitem{Die03}
V.~Dierolf, C.~Sandmann, S.~Kim, V.~Gopalan, and K.~Polgar,
J.~Appl.~Phys. \textbf{93},2295 (2003).

\bibitem{Foe99}
M.~Foeth, A.~Sfera, P.~Stadelmann, and P.-A. Buffat, J.~Electron.\
Microsc.\ \textbf{48}, 717 (1999).

\bibitem{Cho05b}
Y.~Cho in Ferroelectric Thin Films, Topics Appl.\ Phys.\
\textbf{98}, 105 (2005) Springer (Berlin Heidelberg)

\bibitem{New}
R.~E.~Newnham, {\it Properties of Materials} (Oxford University
Press, New York, 2005)

\bibitem{Alexe}
M.~Alexe and A.~Gruverman, eds., {\it Nanoscale Characterisation of
Ferroelectric Materials} (Springer, Berlin; New York, 2004) 1st ed.

\bibitem{Kol95}
O.~Kolosov, A.~Gruverman, J.~Hatano, K.~Takahashi, and H.~Tokumoto,
Phys.\ Rev.\ Lett. \textbf{74}, 4309 (1995).

\bibitem{Jun06}
T.~Jungk, A.~Hoffmann, and E.~Soergel, Appl.\ Phys.\ Lett.\
\textbf{89}, 163507 (2006).

\bibitem{Jun07}
T.~Jungk, A.~Hoffmann, and E.~Soergel,
J. of Microsc. accepted (2007).

\bibitem{All70}
H.~Allik and T.~J.~R.~Hughes, International Journal for Numerical
Methods in Engineering \textbf{2}, 151 (1970).

\bibitem{Tia07}
L.~Tian, P.~Capek, S.~Choudhury, E.~A.~Eliseev, A.~N.~Morozovska,
V.~Dierolf, L.~Chen, S.~Kalinin, and V.~Gopalan, to be published

\bibitem{Greiner}
W.~Greiner, Klassische Elektrodynamik. Harri Deutsch Verlag,
Frankfurt am Main, (2002).

\bibitem{Jun07a}
T.~Jungk, A.~Hoffmann, and E.~Soergel, cond-mat/0602137

\bibitem{Rod05}
B.~J.~Rodriguez, R.~J.~Nemanich, A.~Kingon, A.~Gruverman,
S.~V.~Kalinin, K.~Terabe, X.~Y.~Liu, and K.~Kitamura,
Appl.\ Phys.\ Lett.\ \textbf{86}, 012906 (2005).

\end{thebibliography}
\end{document}